\documentclass[12pt]{article}
\usepackage{amssymb}
\usepackage{euscript} 
\def\ln{{\rm ln}}
\def\a{\begin{eqnarray}}
\def\b{\end{eqnarray}}
\def\0{\nonumber}
\def\ba{\begin{array}}
\def\ea{\end{array}}

\def\um{\frac{1}{2}}
\def\A{{\cal A}}

\def\Tr{{\rm Tr}}

\newcommand{\de}{\mathrm{d}}

\def\q{{\widetilde{\cal Q}}}

\def\al{{\alpha}}
\def\lm{{\lambda}}

\def\cm{{\cal M}}

\renewcommand{\theequation}{\thesection.\arabic{equation}}

\newlength{\extraspace}
\newlength{\extraspaces}
\newcounter{dummy}
\newcommand{\ai}{
\addtocounter{equation}{1}
\setcounter{dummy}{\value{equation}}
\setcounter{equation}{0}
\renewcommand{\theequation}{\thesection.\arabic{dummy}\alph{equation}}
\begin{eqnarray}
\addtolength{\abovedisplayskip}{\extraspaces}
\addtolength{\belowdisplayskip}{\extraspaces}
\addtolength{\abovedisplayshortskip}{\extraspace}
\addtolength{\belowdisplayshortskip}{\extraspace}}
\newcommand{\bj}{
\end{eqnarray}
\setcounter{equation}{\value{dummy}}
\renewcommand{\theequation}{\thesection.\arabic{equation}}}
\def\d{{\partial}}

\def\ddt2{{{\d}\over{\d t_2}}}

\newcommand{\bac}{\begin{array}{c}}
\newcommand{\bacc}{\begin{array}{cc}}
\newcommand{\baccc}{\begin{array}{ccc}}
\newcommand{\barcl}{\begin{array}{rcl}}
\newcommand{\bacccc}{\begin{array}{cccc}}
\newcommand{\baccccc}{\begin{array}{ccccc}}
\newcommand{\baccccccc}{\begin{array}{ccccccc}}
\newcommand{\barclcrcl}{\begin{array}{rclcrcl}}
\newcommand{\bacl}{\begin{array}{cl}}
\newcommand{\bal}{\begin{array}{l}}
\newcommand{\bacll}{\begin{array}{cll}}
\begin{document}
\begin{flushright}
SISSA-ISAS 84/2005/EP\\
hep-th/0511152
\end{flushright}
\vskip0.5cm
\centerline{\LARGE\bf Conifold geometries, matrix models}
\centerline{\LARGE\bf and quantum solutions.}
\vskip 1cm
\centerline{\large G.Bonelli, L.Bonora, A.Ricco}
\centerline{International School for Advanced Studies (SISSA/ISAS)}
\centerline{Via Beirut 2, 34014 Trieste, Italy}
\centerline{INFN, Sezione di Trieste.  }
\vskip 2cm
\abstract{This paper is a continuation of hepth/0507224 where open 
topological B-models describing D-branes on 2-cycles of local 
Calabi--Yau geometries with conical singularities were studied. After a 
short review, the paper expands in particular on two aspects: 
the gauge fixing problem 
in the reduction to two dimensions and the quantum matrix model solutions.}

\vskip 1cm

\section{Introduction}

Singular Calabi--Yau manifolds represent one of the most interesting 
developments in string compactifications. For instance, the presence of a 
conifold point, \cite{candelas}, in a Calabi-Yau opens new prospects: in 
conjunction with fluxes and branes it may allow for warped compactifications, 
which in turn may create the conditions for moduli stabilization
and for large hierarchies of physical scales. On the other hand
singular Calabi-Yau compactifications with conical singularities seem to 
realize favorable conditions for low energy theory models with realistic 
cosmological features. 
A conifold singularity can be smoothed in two different ways, by means of 
a 2-sphere (resolution) or a 3--sphere (deformation). This leads, from a 
physical point of view, to a geometric transition that establishes a duality 
relation between theories defined by the two nonsingular geometries 
(gauge--gravity or open--closed string duality),\cite{Cachazo1,Cachazo2}. 
In summary, conifold singularities are at the 
crossroads of many interesting recent developments in string theory. It is
therefore important to study theories defined on conifolds, i.e. on
singular non compact Calabi-Yau threefolds, as well--defined and (partially)
calculable models to approximate more realistic situations.    

In \cite{BBR}, building on previous literature, we started to elaborate
on an idea that is receiving increasing attention: 
how data describing the geometry of a local 
Calabi--Yau can be encoded, via a topological field theory, in a 
(multi--)matrix 
model and how they can be efficiently calculated. The framework we 
considered was
IIB string theory with spacetime filling D5--branes wrapped around 
resolving two--dimensional cycles. This geometry defines a 4D gauge theory, 
\cite{KW,Dijk1,Dijk2,Dijk3}. On the other hand one can consider 
the open topological B model describing strings on the conifold. 
The latter has been 
shown long ago by Witten to be represented by a six-dimensional holomorphic 
Chern--Simons theory, \cite{Witten}. When reduced to a two-dimensional cycle 
this theory can be shown to reduce to a matrix model. In particular, 
if one wishes to represent a wide class of deformations of the 
complex structure satisfying the Calabi--Yau condition, one must resort to
very general multi--matrix models, \cite{Ferrari}. 
In \cite{BBR} we concentrated on the topological string theory
part of the story, \cite{agan,Diaconescu,Marino,Bilal,neitzke,curto},
in particular on the formal aspects of the reduction 
from the six-dimensional holomorphic Chern--Simons theory to a two--dimensional 
field theory and on the analysis of the matrix model potentials originated from
the Calabi-Yau complex structure deformations. Finally we concentrated
on the subclass of 
matrix models represented by two--matrix models with bilinear coupling
(for a very recent development see \cite{Lech}). In this 
case the functional integral can be explicitly calculated with the method of 
orthogonal polynomials. Using old results we showed how one 
can find explicit solutions by solving the {\it quantum equations of motion} 
and utilizing the recursiveness granted by integrability. All the data turn
out to be encoded in a Riemann surface, which we called 
{\it quantum Riemann surface} in order to distinguish it from the Riemann 
surface of the standard saddle point approach.    

In this paper we would like to return to some topics that were only partially
developed in \cite{BBR}. In particular, in section 2, after a concise review of
the reduction to from CS theory to matrix models, we explain in detail
the gauge fixing problem in this process. Subsequently we return
to the problem of solving two--matrix models with bilinear couplings by means 
of the orthogonal polynomials method via the solution of the quantum equation 
of motion. After a short review of the method in section 3,
our main purpose in section 4 is to clarify the similarities and differences
with other methods, in particular with the semiclassical saddle point method.
We discuss at length the result of \cite{BBR} that the quantum equations of
motion admit in general more vacua than the saddle point method. We interpret 
these additional solutions as `quantum' cycles that have no classical analog. 
Finally in section 5 we give a simple, explicit example of topological open 
string expansion.

\section{Reduction to the brane}

In this section we summarize the reduction of the topological open string 
field theory (B model) to a holomorphic 2-cycle in a local Calabi-Yau threefold 
\cite{BBR}. 
Let us consider a holomorphic $\mathbb{C}^2-$fibre bundle $X \to \Sigma$, where
 $\Sigma$ is a Riemann surface. The space $X$ is obtained as a deformation of 
 the complex structure of the total space of a rank-2 vector bundle $V$ on 
 $\Sigma$. Given an atlas $\{ U_\alpha \}$ on $\Sigma$, the transition 
 functions for $X$ can be written
\begin{eqnarray}
z_{(\alpha)} &=& f_{(\alpha\beta)}(z_{(\beta)}) \nonumber \\
\omega^i_{(\alpha)} &=& M^i_{j(\alpha\beta)}\left(z_{(\beta)}\right)
                      \left[ \omega^j_{(\beta)} + \Psi^j_{(\alpha\beta)}
		      \left(z_{(\beta)},\omega_{(\beta)}\right) \right]\ , 
		      \quad i,j=1,2\0
\end{eqnarray}
where $f_{(\alpha\beta)}$ are the transition functions on the base,
$M^i_{j(\alpha\beta)}$ the transition functions of the vector bundle $V$ 
and $\Psi^j_{(\alpha\beta)}$ are the deformation terms, holomorphic on the 
intersections $(U_\alpha \cap U_\beta) \times \mathbb{C}^2$.

The Calabi-Yau condition on the space $X$, i.e. the existence of a nowhere
vanishing holomorphic top-form $\Omega = \de z \wedge \de w^1 \wedge \de w^2$, 
puts conditions on the vector bundle and on the deformation terms. The 
determinant of the vector bundle has to be equal to the canonical line bundle 
on $\Sigma$ and for the transition functions this means 
$\det M_{(\alpha \beta)} \times f'_{(\alpha \beta)}=1$. For the deformation 
terms we have $\det \left(1+\partial\Psi\right) = 1$, where 
$(1+\partial\Psi)^i_j=\delta^i_j+ \partial_j\Psi^i$.
The solution of this condition can be given in terms of a set of potential 
functions $X_{(\alpha\beta)}$, the \emph{geometric potential}, which generates 
the deformation via 
\begin{eqnarray}
\epsilon_{ij} w^i_{(\alpha\beta)} \de w^j_{(\alpha\beta)} = \epsilon_{ij} 
\omega^i_{(\alpha)} \de \omega^j_{(\alpha)} - \de X_{(\alpha\beta)},\0
\end{eqnarray}
where we define the \emph{singular coordinates} $w^i_{(\alpha\beta)}=
\omega^i_{(\alpha)} + \Psi^i_{(\alpha\beta)}(z^{(\beta)},\omega_{(\beta)})$.

The topological open B-model on $X$ can be obtained from open string field 
theory and reduces \cite{Witten} to the holomorphic Chern-Simons (hCS) theory 
on $X$ for a (0,1)-form connection on a $U(N)$ bundle $E$, where $N$ is the 
number of space-filling B-branes. We will restrict ourselves to the case in which $E$ 
is trivial. The action of hCS is
\begin{eqnarray}
S(\mathcal{A})=\frac{1}{g_s}\int_X {\cal L}, 
\quad\quad
{\cal L}=
\Omega \wedge Tr\left(\frac{1}{2}  \mathcal{A}\wedge
\bar\partial \mathcal{A} + \frac{1}{3} \mathcal{A} \wedge \mathcal{A} \wedge 
\mathcal{A} \right)
\label{hCS}
\end{eqnarray}
where $ \mathcal{A}\in T^{(0,1)}(X)$. The dynamics of B-branes on a 2-cycle 
$\Sigma \subset X$ can be described by reducing the open string field theory 
from the space $X$ to the B-brane world-volume $\Sigma$.

To obtain the reduced action for the \emph{linear geometry} ($\Psi_i \equiv 0$),
first we split the form $\mathcal{A}$ into horizontal and vertical components 
using a reference connection $\Gamma$ on the vector bundle, then we impose 
the independence of the fields on the vertical directions and finally we 
``integrate along the fiber'' using a bilinear structure $K$ on the bundle. 
If the connection $\Gamma$ is the generalized Chern connection for the 
bilinear structure, then the result is independent of the particular 
$(\Gamma, K)$ chosen.

Let us define $\A_{\bar z} = A_{\bar z} - 
A_{\bar k} \Gamma_{\bar z \bar j}^{\bar k} \bar w^{\bar j}$ and 
$\A_{\bar i} = A_{\bar i}$, where $\Gamma$ is a reference connection and 
impose that the components $(A_{\bar z},A_{\bar i})$ are independent on the 
coordinates along ${\mathbb C}^2$, obtaining for the Lagrangian 
\begin{eqnarray}\label{redlag}
L = \frac{1}{2} \Omega \wedge \Tr  \left\{ A_{\bar i} D_{\bar z} A_{\bar j} 
+ A_{\bar i} \Gamma_{\bar z\bar j}^{\bar k}A_{\bar k} \right\}
\de w^{\bar i} \wedge \de \bar z\wedge \de w^{\bar j}
\end{eqnarray}
where $D_{\bar z}$ is the covariant derivative w.r.t. the gauge structure.

Now let us consider a bilinear structure $K$ in $V$, {i.e.} a local section 
$K\in \Gamma(V\otimes \bar V)$, the components $K^{i \bar j}$ being an 
invertible complex matrix at any point. 
The ``integration along the fiber'' is realized contracting the hCS 
(3,3)-form Lagrangian by the two bi-vector fields $k = 
\frac{1}{2} \epsilon_{ij}K^{i \bar l}K^{j \bar k} \frac{\partial}{\partial 
\bar w^l} \frac{\partial}{\partial \bar w^k}$ and $\rho=\frac{1}{2} 
\epsilon^{ij}\frac{\partial}{\partial w^i} \frac{\partial}{\partial w^j}$
\begin{eqnarray}\label{pb'}
{\cal L}_{red} = i_{\rho\wedge k} L = 
                 \um \de z \de {\bar z} (\det K) \epsilon^{\bar{i} \bar{j}} 
                     \Tr \left[ A_{\bar i}D_{\bar z}A_{\bar j} + 
		     A_{\bar i}\Gamma_{\bar z\bar j}^{\bar k}A_{\bar k} \right].
\end{eqnarray}
Defining the field components $\varphi^i=i_{V^i}A\in V$, where 
$V^i=K^{i \bar j}\frac{\partial}{\partial\bar w^j}$, one gets
\begin{eqnarray}\label{pb''}
{\cal L}_{red} = \um \de z \de{\bar z} \Tr\left[\epsilon_{ij} 
\varphi^iD_{\bar z}\varphi^j + (\det K) \varphi^m\varphi^n 
\epsilon^{\bar i\bar j} 
                 \left( K_{m \bar i}\partial_{\bar z}K_{n \bar j}
		 +K_{m \bar i}K_{n \bar k} \Gamma_{\bar z\bar j}^{\bar k}
		  \right)\right]
\end{eqnarray}
where $K_{\bar i j}$ are the components of the inverse bilinear structure, 
that is $K_{\bar i j} K^{j \bar l}=\delta_{\bar i}^{\bar l}$.
In order to have a result which is independent of the trivialization, just
set the reference connection to be the generalized Chern connection of 
the bilinear structure $K$, that is
$\Gamma_{\bar z\bar j}^{\bar k}= K_{\bar j l}\partial_{\bar z}K^{l\bar k}$.
The action for the reduced theory is given by
\begin{eqnarray}
S_{red}=\frac{1}{g_s} \int_\Sigma {\cal L}_{red} = 
\frac{1}{2 g_s}\int_\Sigma \de z \de {\bar z}\Tr\left[ 
\epsilon_{ij}\varphi^i D_{\bar z}\varphi^j \right].\0
\end{eqnarray}

In the \emph{rational case} $\Sigma \simeq \mathbb{P}^1$ with non vanishing 
deformation terms $\Psi_i$, $X$ is a deformation of a vector bundle 
$V \simeq \mathcal{O}_{\mathbb{P}^1}(n) \oplus \mathcal{O}_{\mathbb{P}^1}(-n-2)$
for some $n$.

Let us start with the reduction in  the {\it Abelian $U(1)$ case}.
In this case the cubic term in the hCS Lagrangian is absent and the reduction 
is almost straightforward. In the singular coordinates $(\varphi^1, \varphi^2)$ 
we obtain that
\begin{equation}
{\cal L}_{red}=\um\epsilon_{ij}\varphi^i \partial_{\bar z} \varphi^j dz 
\de {\bar z}\0
\label{pbsing}\end{equation}
in both charts of the standard atlas $\{ U_N, U_S \}$ of $\mathbb{P}^1$. 
The potential term $X$ gives the deformation of the action due to the 
deformation of the complex structure. Passing to the non singular 
coordinates $(\phi_1, \phi_2)$, one obtains
\begin{equation}
S_{red} = \frac{1}{2\,g_s}
          \left[ \int_{\mathbb{P}^1}  \de z \de \bar z \epsilon_{ij}\phi^i 
	  \partial_{\bar z} \phi^j 
          + \oint \frac{\de z}{2 \pi i} X(z,\phi) \right]
\end{equation}
where $\oint$ is a contour integral along the equator. Therefore, the reduced 
theory gives a $b$--$c$ ($\beta$--$\gamma$) system on the two hemispheres 
with a junction interaction along the equator.

The {\it non--Abelian case} is a bit more complicated than the Abelian one 
because of the tensoring with the (trivial) gauge bundle. 
This promotes the vector bundle sections to matrices and therefore it is 
not immediate how to unambiguously define the potential function $X$ in 
the general case. 
The easiest way to avoid matrix ordering prescriptions is to 
restrict to the case in which 
$X(z,\omega)$ does not depend, say, on $\omega^2$.
Defining $B := \omega^1 \Psi^2 + X$, one obtains $\Psi^1 = 0$, $\Psi^2 = 
\partial_{\omega^1} B$ and the reduced action
\begin{eqnarray}
S\equiv S_{red} = \frac{1}{g_s} \left[\int_{\mathbb{P}^1} -\Tr(\phi^2 D_{\bar z} 
\phi^1) \de z \de \bar z + \oint \Tr B(z,\phi^1) \de z\right]\label{Sred}
\end{eqnarray}
 
\subsection{Gauge fixing}

In order to calculate its partition function,
let us now discuss the gauge fixing of the theory.
The following discussion is a refinement of the derivation given in 
\cite{Dijk1}. Our starting action is (\ref{Sred})
and we follow the standard BRST quantization (see for example \cite{HT}).

The BRST invariance in the minimal sector is 
$$
sA_{\bar z}=-(Dc)_{\bar z},\quad
s\phi^i=[c,\phi^i],\quad
sc=\frac{1}{2}[c,c]
$$
while we add a further non minimal one to implement the gauge fixing with
$$
s\bar c=b,\quad sb=0.
$$

The gauge fixed action is obtained by adding to $S$ a gauge fixing term
$$
S_{gf}=S+s\Psi,\quad {\rm where}\quad
\Psi=\frac{1}{g_s}\int_{P^1}{\rm Tr}\bar c \partial_zA_{\bar z}
$$
which implements a holomorphic version of the Lorentz gauge.
Actually we have
$$s\Psi=\frac{1}{g_s}\int_{P^1}{\rm Tr}\left[b\partial_z A_{\bar z}- 
\partial_z\bar c(Dc)_{\bar z}\right].$$

Our partition function is then the functional integral
$$
Z_B=\int{\cal D}\left[\phi^i,A_{\bar z},c,\bar c,b\right]
e^{-S_{gf}}
$$
The calculation can proceed as follows.
Let us first integrate along the gauge connection $A_{\bar z}$ which 
enters linearly the gauge fixed action
and find
$$
Z_B=\int{\cal D}\left[\phi^i,c,\bar c,b\right]
e^{-\frac{1}{g_s}\left[-\int_{P^1}{\rm Tr}\phi^2\partial_{\bar z}\phi^1
- \partial_z\bar c\partial_{\bar z}c
+\oint{\rm Tr}B(z,\phi^1)\right]
}
\delta\left\{
\partial_z b + [\partial_z\bar c, c] +[\phi^1,\phi^2]
\right\}
$$
Now we integrate along the field $b$. By solving the constraint we obtain
$$
Z_B=\int{\cal D}\left[\phi^i,c,\bar c\right]
e^{-\frac{1}{g_s}\left[-\int_{P^1}{\rm Tr}\phi^2\partial_{\bar z}\phi^1
- \partial_z\bar c\partial_{\bar z}c
+\oint{\rm Tr}B(z,\phi^1)\right]
}\frac{1}{{\rm det}' \partial_z}
$$
where ${\rm det}'$ is the relevant functional determinant with the exclusion of
the zero modes. Then we integrate along the $(c,\bar c)$ ghosts and get
$$
Z_B=\int{\cal D}\left[\phi^i\right]
e^{-\frac{1}{g_s}\left[-\int_{P^1}{\rm Tr}\phi^2\partial_{\bar z}\phi^1
+\oint{\rm Tr}B(z,\phi^1)\right]
}\frac{
{\rm det}' \partial_z\partial_{\bar z}
}{{\rm det}' \partial_z}
$$
Finally, since the geometric potential $B$ does not depend on $\phi^2$, 
we can also integrate along this 
variable and obtain
$$
Z_B=\int{\cal D}\left[\phi^1\right]
e^{-\frac{1}{g_s}\left[\oint{\rm Tr}B(z,\phi^1)\right]}
\delta(\partial_{\bar z}\phi^1)
\frac{
{\rm det}' \partial_z\partial_{\bar z}
}{{\rm det}' \partial_z}
$$
The delta function constrains the field $\phi^1$ to span the 
$\partial_{\bar z}$-zero modes and once it is solved it 
produces a further $\left({\rm det}' \partial_{\bar z}\right)^{-1}$ 
multiplicative term that cancel the other determinants.
Therefore, all in all we get 
$$
Z_B=\int_{{\rm Ker}\partial_{\bar z}} d\phi^1
e^{-\frac{1}{g_s}\left[\oint{\rm Tr}B(z,\phi^1)\right]}.
$$
Lastly we can expand $\phi^1=\sum_{i=0}^n X_i\xi_i$ along the basis 
$\xi_i(z)\sim z^i$
of ${\rm Ker}\partial_{\bar z}$ with $N\times N$ matrix coefficients $X_i$.
Finally we find the multi-matrix integral
\a
Z_B=\int \prod_{i=0}^n d X_i
e^{-\frac{1}{g_s}{\cal W}\left(X_0,\dots,X_n\right)}\label{functint}
\b
where we defined
\a
{\cal W}\left(X_0,\dots,X_n\right)
=\oint{\rm Tr}B(z,\sum_i X_i z^i)\label{B}
\b

This is the result of our gauge fixing procedure which covers the details 
needed to complete the derivation 
given in \cite{BBR} and confirms the conjecture in \cite{Ferrari}.

\section{General properties of two--matrix models.}

\setcounter{equation}{0}
\setcounter{footnote}{0}

The second part of this paper is devoted to solving some of the matrix models
introduced above, eq.(\ref{functint}), the two--matrix models with bilinear
coupling. The sequel is a short review 
of \cite{BBR} on this subject containing some additional remarks and 
complements. The main point we
insist on is that for these models there is the possibility to solve the 
quantum problem exactly. That is, we perform the path integral exactly 
and determine all the (quantum) solutions. The exact solvability of these
two--matrix model is a well--known fact, but its consequences have not
yet been completely explored. As we will see, not all these 
solutions have a classical analog and, thus, they represent genuine new 
quantum solutions. 
To this purpose the method of orthogonal polynomials turns out to be
particularly fit. This method allows one to explicitly perform the path 
integration, so that one is left with quantum equations of motion and the flow
equations of an integrable linear systems. The latter in particular
uncover the integrable nature of two--matrix models, which stems from the Toda 
lattice hierarchy underlying all of them. 
Our approach for solving two--matrix models consists in solving the 
quantum equations of motion and, then, using the
recursiveness intrinsic to integrability (the flow equations), in finding 
explicit expressions for the correlators. An alternative method 
is based on the W constraints on the functional integral.
We do not use it here, but one can find definitions, applications and 
comparisons with the other methods in \cite{BX1,BCX}. 
For general references on matrix models, see the bibliography in 
\cite{eynard,BBR}.

Let us start with a synthetic review of the approach based on orthogonal 
polynomials.  
The model of two Hermitian $N \times N$ matrices $M_1$ and $M_2$ with bilinear
coupling is defined by the 
partition function 
\a
Z_N(t,c)=\int dM_1dM_2 e^{tr W},\quad\quad
W=V_1 + V_2 + c M_1 M_2\label{Zo}
\b
with potentials
\a
V_{\al}=\sum_{r=1}^{p_\al} \bar t_{\al,r}M_{\al}^r\,\qquad \al=1,2.\label{V}
\b
where $p_\al$ are finite numbers.  
We denote by ${\cal M}_{p_1,p_2}$ the corresponding two--matrix model. 
With reference to eq.(\ref{B}), this model descends from the geometric potential
$B$ defined by
\begin{eqnarray}
B(z, \omega) = \frac{1}{z} \left[V_1(\omega)+V_2\left(\frac{\omega}{z}\right)\right]
+\frac{c}{2z^2}\omega^2\label{2mpot}
\end{eqnarray}

We are interested in computing correlation functions (CF's) of the operators
\a
\tau_k=tr M_1^k,\qquad \sigma_k=tr M_2^k ,\qquad
\forall k,\0
\b
For this reason we complete the above model by replacing
(\ref{V}) with the more general potentials
\a
V_\al = \sum_{r=1}^\infty  t_{\al,r} M_\al^r, \qquad \al =1,2\label{Vgen}
\b
where $t_{\al , r} \equiv \bar t_{\al,r}$ for $r\leq p_\al$. The CF's are
defined by
\a
< \tau_{r_1}\ldots \tau_{r_n}\sigma_{s_1}\ldots \sigma_{s_m}> =
\frac {\d^{n+m} }{\d t_{1,r_1}\ldots \d t_{1,r_n}\d t_{2,s_1}\ldots 
\d t_{2,s_m}} \ln Z_N(t,g)\label{CF}
\b
where, in the RHS, all the $t_{\al,r}$ are set equal to $\bar t_{\al,r}$
for $r\leq p_\al$ and the remaining are set to zero.   
From now on we will not distinguish between $t_{\al,r}$ and $\bar t_{\al,r}$
and use throughout only $t_{\al,r}$. We hope the context will always make 
clear what we are referring to. 

We recall that the ordinary procedure to calculate the partition function 
consists of three steps \cite{BIZ},\cite{IZ2},\cite{M}:
$(i)$ one integrates out the angular part so that only the
integrations over the eigenvalues are left;
$(ii)$ one introduces the orthogonal monic polynomials
\a
\xi_n(\lambda_1)=\lambda_1^n+\hbox{lower powers},\qquad\qquad
\eta_n(\lambda_2)=\lambda_2^n+\hbox{lower powers},\quad\quad n=0,1,2,\ldots\0
\b
which satisfy the orthogonality relations
\a
\int  d\lambda_1d\lambda_2\xi_n(\lambda_1)
e^{V_1(\lm_1)+V_2(\lm_2)+c\lm_1\lm_2}
\eta_m(\lambda_2)=h_n(t,c)\delta_{nm}\label{orth1}
\b
$(iii)$, using the orthogonality relation (\ref{orth1}) and the properties
of the Vandermonde determinants, one can easily
calculate the partition function
\a
Z_N(t,c)={\rm const}~N!\prod_{i=0}^{N-1}h_i\label{parti1}
\b
whereby we see that knowing the partition function amounts to knowing
the coefficients $h_n(t,c)$. The crucial point is that the information 
concerning the latter can be encoded in the flow equations of the
Toda lattice hierarchy and the quantum equations of motion. Before 
coming to this, let us introduce some convenient notations. 
In the sequel we will meet infinite matrices 
$M_{ij}$ with $0\leq i,j <\infty$. For any such matrix $M$, we define
\a
\cm=H^{-1} MH,\qquad H_{ij}=h_i\delta_{ij},\qquad \widetilde M_{ij} = M_{ji}\0
\b
We represent such matrices in the lower right quadrant of the $(i,j)$ plane.
They all have a band structure, with nonzero elements belonging
to a band of lines parallel to the main descending diagonal. 
We will write $M\in [m,n]$, if all its non--zero lines are between
the $m$--th and the $n$--th ones, setting $m=0$ for the main diagonal.
We refer to such types of band matrices as Jacobi matrices.
Moreover $M_+$ will denote the upper triangular
part of $M$ (including the main diagonal), while $M_-=M-M_+$.  

Let us come now to the  {\it quantum equations of motion}. They are written as
\a
P^\circ(1)+ V_1'(Q(1))+c Q(2)=0,\qquad\quad
c Q(1)+V_2'(Q(2))+\widetilde{\cal P}^\circ(2)=0,\label{coupling}
\b
In these equations $Q(1),Q(2), P^\circ(1),P^\circ(2)$ are infinite Jacobi 
matrices. They represent the multiplication by $\lambda_1,\lambda_2$ and the
derivative by the same parameters, respectively, in the basis of monic
polynomials.
Eqs.(\ref{coupling}) can be considered the quantum analog of the classical 
equations of motion.
The difference with the classical equations of motion of the original matrix
model is that, instead of the $N \times N$ matrices $M_1$ and $M_2$, here we 
have infinite $Q(1)$ and $Q(2)$ matrices together with the quantum deformation 
terms given by $P^\circ(1)$ and $\widetilde {\cal P}^\circ(2)$, respectively.
From the coupling conditions it follows at once that
\a
Q(\al)\in[-m_{\al}, n_{\al}],\qquad \al=1,2\0
\b
where
\a
m_1=p_2-1, \qquad\quad m_2=1 \qquad\quad\quad
n_1=1, \qquad\quad n_2=p_1-1\0
\b
where $p_\al$, $\al =1,2$ is the highest order of
the potential $V_\al$ (see (\ref{V})).

The {\it flow equations} of the Toda lattice hierarchy are 
\ai
&&{\partial\over{\partial t_{\al,k}}}Q(1)=[Q(1),~~Q^k(\al)_-],\qquad \alpha
=1,2\label{CC12}\\
&&{\partial\over{\partial t_{\al,k}}}Q(2)=[Q^k(\al)_+,~~Q(2)]
\label{CC22} 
\bj

Finally one must use the reconstruction formula for the partition function  
\a
{\d \over \d t_{\al, r}} \ln Z_N(t,c) = \sum_{i=0}^{N-1}
\Big(Q^r(\al)\Big)_{ii}, \quad\quad
\al = 1,2 \label{ddZ}
\b
It is evident that, by using the equations (\ref{CC12},\ref{CC22}) above 
we can express all the derivatives of $Z_N$ in terms of the elements of the 
$Q$ matrices. For example
\a
{\d^2\over{\d t_{1,1}\d t_{\al,r}}}
\ln Z_N(t,c)=\Bigl(Q^r(\al)\Bigl)_{N,N-1},\qquad \al = 1,2\label{parti3}
\b
and so on. We recall that the derivatives of $F(N,t,c) = \ln Z_N(t,c)$ at
$t_{\alpha,r}=\bar t_{\alpha,r}$ are 
nothing but the correlation functions of the model.

To end this section, we collect a few formulas we will need later on.
First, we will be using the following parametrization of 
the Jacobi matrices
\a
Q(1)=I_++\sum_i \sum_{l=0}^{m_1} a_l(i)E_{i,i-l}, \qquad\qquad\qquad
\q(2)=I_++\sum_i \sum_{l=0}^{m_2} b_l(i)E_{i,i-l}\label{jacobi1}
\b
where $I_+ = \sum_{i=0}E_{i,i+1}$ and $(E_{i,j})_{k,l}=
\delta_{i,k}\delta_{j,l}$. One can immediately see that
\a
\Bigl(Q_+(1)\Bigl)_{ij}=\delta_{j,i+1}+a_0(i)\delta_{i,j},\qquad
\Bigl(Q_-(2)\Bigl)_{ij}=R(i)\delta_{j,i-1}\label{jacobi2}
\b
where $R({i+1}) \equiv h_{i+1}/h_i$.
As a consequence of this parametrization, eq.(\ref{parti3}) gives in
particular the two important relations
\a
{\d^2\over{\d t^2_{1,1}}}
F(N,t,c)=a_1(N), \quad\quad 
\frac {\partial^2}{\d t_{1,1} \d t_{2,1}}F(N,t,c) = R(N)\label{FR}
\b
 
To complete this summary, we should mention the W--constraints method. 
The latter are constraints on the partition function which take the form of 
a nice algebraic structure, see \cite{BX1,BCX} for instance. They are 
obtained by putting together quantum equations of motion and flow equations. 
W--constraints (which are also called loop equations or Schwinger--Dyson
equations) can be used to solve matrix models, but such a 
procedure is less efficient than the one used here.
 
Finally we must recall that the CF's we compute are genus expanded. 
The genus expansion is strictly connected with the homogeneity properties 
of the CF's. The contribution
pertinent to any genus is a homogeneous function of the couplings (and $N$)
with respect to appropriate degrees assigned to all the involved quantities,
see \cite{BBR}. 
In particular we expect the partition function to have, in the large $N$ limit,
the expansion
\a
F= \sum_{g=0}^\infty N^{2-2g} F_g \label{genusexp}
\b
where $g$ is the genus. Such expectation, based on a path integral analysis, 
remains true in our setup due to the fact that the homogeneity properties 
carry over to the Toda lattice hierarchy.  
In (\ref{genusexp}) $F_g$ is interpreted as the result of summing 
the open string partition function at fixed $g$ over all boundaries. 
Therefore it is a closed string quantity and the correlators obtained from it
are closed string correlators, i.e. correlators of the geometrically
dual closed strings relative to the deformed geometry. From matrix models
it is however possible to obtain also genuine open string quantities.
To this end we must change expansion. We introduce the string coupling $g_s$
and rescale all couplings in $W$, so as to extract an overall
factor $1/g_s$ out of them. Now the open string expansion is postulated to be
\a
F= \sum_{h=0}^\infty g_s^{2g-2+h} N^h F_{g,h}\label{Fgh}
\b  
$F_{g,h}$ refers to the contribution
from a world--sheet of genus $g$ with $h$ boundaries.

\section{Solving two--matrix models}

\setcounter{equation}{0}
\setcounter{footnote}{0}

Our procedure to solve two--matrix models consists in solving the quantum 
equations of motion. This allows us to determine the 
`lattice fields' $a_i(n), b_i(n)$ and $R(n)$. Once these are known we can compute 
all the correlation functions starting from (\ref{ddZ}) by repeated use of 
eqs.(\ref{CC12},\ref{CC22}), which form the flows of the
Toda lattice hierarchy. As for the free energy $F(t,N,c)$ itself, see for 
instance (\cite{BX1}). In the following we describe some explicit examples 
of this method. In reality we will concentrate on solving the equations of
motion, since the calculus of correlators is of algorithmic nature and,
therefore, not particularly interesting; in any case, it has already 
been illustrated in a number of examples, \cite{BCX,BX1,BBR}.
The equations of motion are definitely more interesting, because some aspects
of them have not been stressed enough or ignored in the existing literature.
Therefore our purpose is to use some (simple) examples first of all in order 
to exemplify our method and compare it with others, in particular
with the saddle point method; second, in order to single out the novelties
with respect to the existing literature. Our exposition will be very close
to \cite{BBR}, but with more examples and more details. For other approaches
to matrix models see, for instance, \cite{eynard} and references therein.

\subsection{The Gaussian model}.

The bi-Gaussian model ${\cal M}_{2,2}$ was fully solved in \cite{BCX}. 
It is a very simple model, but we further simplify it by considering the
decoupling limit $c\to 0$ and keeping only half of couplings, say $t_k=t_{1,k}$.
This operation makes sense and leads to the Gaussian one--matrix model 
${\cal M}_{2}$.
It allows us to make an explicit comparison of our method and the 
traditional one based on eigenvalue density and resolvent. 

The quantum equations of motion of the simplified model are, see \cite{BCX},
\a
t_1 + 2 t_2a_0(n)=0,\quad\quad n+ 2 t_2 a_1(n)=0\label{M20}
\b
We can set $t_1=0$, because the linear term in the potential can always be 
gotten rid of with a matrix redefinition. Therefore we have $a_0=0$ and 
$a_1=-n/(2t_2)=g_s n/N$. In the last equality we have introduced the string
coupling $g_s = - N/(2 t_2)$, thus making contact with the notation
of \cite{Marino} where $1/g_s$ is factored out in front of the potential in the
path integral. Since the latter reference contains a compact
review of the Gaussian one--matrix model with the saddle point method, we
refer to it for a comparison. 
In the large $N$ limit, $n/N$ becomes the continuous variable
$x$, and so, in this limit, $a_1(x)=g_s x$. Now we can easily compute all 
the correlators. For a comparison with \cite{Marino} we need the one--point
functions in genus 0. Following \cite{BX1} and section 5.5 of \cite{BCX}, we 
can compute
\a
&&<\tau_{2k+1}> =0\0\\  
&&<\tau_{2k}>=\int_0^x L^{2k}_{(0)}(y) \,dy= 
\int_0^x \left( \begin{array}{c} 
2k
\\ k 
\end{array}\right) g_s^k 
\frac {y^{k+1}}{k+1} dy \label{1pt}
\b
where $L=\zeta +a_1(x) \zeta^{-1}$ is the genus 0 continuous version of
$Q(1)$, and the label $_{(0)}$ denotes the coefficient of $\zeta^0$. 
In \cite{Marino} the resolvent is defined as
\a
\omega= \frac 1N \sum_{k=0}^\infty \frac {\tau_k}{p^{k+1}}\label{omega}
\b
Therefore in the genus 0 case we have\footnote{In the continuous limit 
the couplings get renormalized, 
$t_k\to \tilde t_k=t_k/N$ and, due to the expansion (\ref{genusexp}),
we define $<\tau_k>_0= \frac{\d F_0}{\d \tilde t_k}$.} 
\a
\omega_0\!&=&\, g_s \sum_{k=0}^\infty \frac {<\tau_k>_0}{p^{k+1}}
=g_s\sum_{k=0}^\infty
\frac {(2k)!}{k!(k+1)!} g_s^{2k}\frac{x^{2k+1}}{p^{k+1}}\0\\
&=&\, \frac 1u \left(1-\sqrt{1-u^2}\right) = \frac 1{2t}
\left(p-\sqrt{p^2-4t}\right) \label{omega0}
\b
which is the result of \cite{Marino} for the resolvent, provided we 
set $t=xg_s$ and $u=\frac {2t}p$. From this we can reconstruct the eigenvalue
density
\a
\rho(\lambda)= \frac 1{2\pi t}\sqrt{4 t -\lambda^2}\0
\b
In the saddle point method this leads to introducing an auxiliary hyperelliptic 
Riemann surface 
\a
y^2= p^2-4t\label{hyper}
\b
where all data of the model are encoded.

As we see from this elementary example, starting from our formulas above
we can reconstruct all the formulas of the saddle point method. In this case
there is no difference between the two methods.

\subsection{The cubic model}

The full ${\cal M}_{3,2}$ model has been discussed at length in \cite{BCX}
and, in particular, in \cite{BBR}. Here we would like to consider
its decoupling limit $c=0$ and single out the cubic potential 
part, which amounts to considering the cubic one--matrix model  ${\cal M}_{3}$.
In the genus 0 limit this model is described by the discrete algebraic equations 
\a
&& a_0^3 + \frac {t_2}{t_3} a_0^2+ \frac 29 \left(\frac{t_2}{t_3}\right)^2 a_0  
- \frac n{3t_3}=0\label{cubica0}\\
&& a_1 = -\frac 12  a_0^2 - \frac 13 \frac {t_2}{t_3}  a_0\label{a1a0}
\b
where, for simplicity and without loss of generality, we have set $t_1=0$.
One can extract from these equations $a_0$ and $a_1$ and calculate all the
correlators with the algorithm described in the previous section. Here we are 
not interested in this, but rather in analyzing eq.(\ref{cubica0}) and its
solutions.

In the large $N$ limit we shift to $x= \frac nN$ and, in order
to make contact with section 4 of \cite{BIZ} for a comparison,
we simplify a bit further our notation setting
$t_2=-\frac N2$ and $t_3=-Ng$, where $g$ is the 
cubic coupling constant there. 
Moreover we denote $z=3g a_0$. Then eq.(\ref{cubica0}) becomes
\a
18 g^2 x + z(1+z)(1+2z)=0\label{cubicf}
\b
This can be solved exactly for $z$ and gives the three solutions
\a
&&z_1= -\frac 12 + \frac 1{2 I(x)} + \frac {I(x)}6\label{z1}\\
&&z_2= -\frac 12 + \frac {1+i\sqrt{3}}{4 I(x)} + \frac {1-i\sqrt{3}I(x)}{12}
\label{z2}\\
&&z_3= -\frac 12 + \frac {1-i\sqrt{3}}{4 I(x)} + \frac {1+i\sqrt{3}I(x)}{12}
\label{z3}
\b
where
\a
I(x)= 3^{1/3}\left(-324 g^2 x +\sqrt{3} \sqrt{-1+34992 g^4 x^2}\right)^{1/3}
\label{I(x)}
\b
From these we can extract three solutions for $a_0$ and, consequently, for
$a_1$. For small $x$ the three solutions can be expanded as follows
\a
&& z_1= -18 g^2 x - 972 g^4 x^2- 93312 g^6 x^3 - 11022480 g^8 x^4 +
O(x^5)\label{f1}\\
&& z_2= -1 -18g^2 x +972 g^4 x^2 -93312 g^6 x^3 + 11022480 g^8 x^4
+ O(x^5) \label{f2}\\
&& z_3 = -\frac 12 + 36 g^2 x + 186624 g^6 x^3 +O(x^5).\label{f3}
\b

The best way to analyze these solutions is to  notice that they represent
a plane curve in the complex $z,x$ plane. It is a genus 0 Riemann surface 
with punctures at $x=0$ and $x=\infty$, made of 
three sheets joined through cuts running from $z=- 1/(\sqrt{3} 108 g^2)$
to $z= 1/(108\sqrt{3}g^2)$.
The solutions (\ref{f1},\ref{f2},\ref{f3}) correspond to the values $z$ takes
near $x=0$, away from the cuts. In order to pass from one solution to
another we have to cross the cuts. We call the Riemann surface so constructed
the {\it quantum Riemann surface associated to the model}\footnote{In the
example of the previous subsection the quantum Riemann surface was just 
one point.}.
This Riemann surface picture is the clue to 
understanding the solutions with multiple brane configurations, as was 
explained in \cite{BBR}.

Let us analyze the meaning of these three solutions.
To this end it is useful to make a comparison with \cite{BIZ}: we see 
that the first solution corresponds to the unique solution found there, which 
corresponds to the minimum of the classical potential (see
below). 
In fact the correspondence with \cite{BIZ} can be made more precise: one can
easily verify that eqs.(46) there are nothing but 
eqs.(\ref{cubica0},\ref{a1a0}), provided we make the identifications:
$a+b=a_0$ and $(b-a)^2= 4 a_1$ and the rescalings $a_0\to \sqrt{x} a_0,
a_1\to x a_1$ and $g\to g/\sqrt{x}$. 
In \cite{BIZ} the interval $(2a,2b)$ represents a cut in the eigenvalue
$\lambda$ plane. Also in \cite{BIZ}, as in the previous subsection, we 
have therefore an auxiliary Riemann surface. The latter is not to be confused
with the quantum Riemann surface defined above, although the two are 
related. 

Let us now discuss the correspondence between our three solutions 
and the classical extrema of the potential. The classical potential for 
the continuous eigenvalue function $\lambda(x)$ 
(which is $\lambda_n/\sqrt{N}$ in the large $N$ limit), is 
$V_{cl}=\frac 12 \lambda^2 +g \lambda^3$. It has extrema at $\lambda =0$ and 
$\lambda= -1/3g$. To find the classical limit in our quantum approach instead, 
we rescale $t_k$, $k=2,3$ as: $t_k\to t_k/\hbar$, and take $\hbar\to 0$. This
amounts to dropping the last term in eq.(\ref{cubica0}). 
The extrema are 
three, $z=0,-1, -1/2$, which corresponds to $a_0=0, -1/3g, -1/6g$,  not two as 
in the classical case. $z=0$ corresponds to the minimum of the potential, 
$z=-1$ corresponds to the 
maximum, while $z=-1/2$ to the flex. The latter solution does not have a
classical analog. 
  
\begin{quote}
\footnotesize

{\bf Discussion}. Before taking seriously this new quantum solution, we have 
to check whether 
our method of solving the quantum problem has any flaw. Our method consists
in solving the quantum equations of motion (\ref{coupling}). One might wonder
whether there are other independent conditions beside (\ref{coupling}) that 
one must impose on the solutions. Eqs.(\ref{coupling}) are obtained from
the identity
\a
\int  d\lambda_1d\lambda_2\, \frac {\d}{\d \lambda_1}\left(\xi_n(\lambda_1)
e^{V_1(\lm_1)+V_2(\lm_2)+c\lm_1\lm_2}
\eta_m(\lambda_2)\right)=0\label{qEOM}
\b
and a similar equation for the derivative with respect to $\lambda_2$. Of course
one has also
\a
\int  d\lambda_1d\lambda_2\, \frac {\d}{\d \lambda_1}\left(\lambda_1^k \xi_n(\lambda_1)
e^{V_1(\lm_1)+V_2(\lm_2)+c\lm_1\lm_2}
\eta_m(\lambda_2)\right)=0\label{qEOMk}
\b
for any $k$. The question is whether these equations imply additional 
constraints on the solutions of the quantum equations of motion. However
one can easily see that eq.(\ref{qEOMk}) can be written as
\a
Q(1)^k\left(P^\circ(1)+ V_1'(Q(1))+c Q(2)\right)=0\label{couplingk}
\b
Therefore a solution to eqs.(\ref{coupling}) is automatically a solution to
(\ref{couplingk}). Thus these additional equations (as well as all those 
that can be obtained by inserting, instead of $\lambda_1^k$, generic monomials 
of $\lambda_1$ and $\lambda_2$ in (\ref{qEOMk})), cannot further constrain 
the solutions to (\ref{coupling}). By the way, the W--constraints are obtained
precisely by taking the trace of all the expressions like (\ref{couplingk})
and expressing them in terms of derivative of $\ln Z$. Therefore the three
quantum solutions we have found above are also solutions to the W--constraints
(or loop equations). Finally, if we consider higher order derivatives 
with respect to $\lambda_1$ in (\ref{qEOM}) instead of the first order one, it
is easy to see that we do not get any additional constraints either.
\end{quote}

This result is somewhat puzzling, but we should remember
that the saddle point method is semiclassical: one cannot exclude that
the quantum problem admits solution without classical analog. This is
precisely what happens in the present case. One can phrase it also by saying 
that, in general, the large $N$ limit and the $\hbar\to 0$ limit do not 
commute.

The next question is: what is the meaning of the third solution, $z=-1/2$?
Let us recall what the other two solutions at $z=0$ and $z=-1$ mean. On the
basis of the discussion in section 2, we know that they represent two 
Riemann spheres located at the minimum and the maximum of the potential.
They replace the continuous family of ${\mathbb P}^1$
which characterizes the conifold geometry before the deformation $W$ is 
introduced. This is the interpretation on the basis of classical geometry. 
What we learn now is that solving the quantum problem we obtain a third
solution, which we can interpret as a {\it quantum } ${\mathbb P}^1$, 
located at the flex of the potential. This is a pure quantum geometry effect.

Before we end this section we would like to make a few remarks. 
First we notice that the classical extrema are characterized by 
the fact that $a_1=0$,  while the pure quantum
solution corresponds to a non-vanishing $a_1$. Moreover, after setting
$a_1=0$ we get for $a_0$ an equation that coincides with the classical 
eigenvalue equation.
From this simple example we learn three important pieces of informations.

\begin{itemize}  

\item The number of solutions of the quantum problem (i.e. the number of 
solutions to 
eq.(\ref{cubicf})) is in general larger than the number of the extrema 
of the classical 
potential.

\item The field $a_0$ can be regarded as the quantum version of the classical 
eigenvalue function.

\item The classical extrema are obtained by neglecting the $n$ term in 
eq.(\ref{cubica0}) and setting $a_1=0$. 

\end{itemize}

These conclusion are valid in general, except for the fact that,
the condition $a_1=0$ in the last remark must be replaced 
by the fields $a_1,a_2,..., b_1,b_2,...,$ being set to zero in the general 
case.

\subsection{The ${\cal M}_{3,3}$ model}

We study the model in the case $t_1=s_1=0$ and limit ourselves to writing down
the genus 0 quantum equations of motion:
\a
&&3t_3 c a_0^2 +2t_2 c a_0 -36 s_3t_3b_0R + c^2 b_0 -12s_2t_3R=0\0\\
&&3 s_3c b_0^2 +2s_2cb_0 -36s_3t_3a_0R -12s_3t_2R + a_0c^2=0\0\\
&&nc +Rc^2 -18 s_3t_3R^2 -36s_3t_3a_0b_0R -12 s_2t_3a_0R -12 t_2s_3b_0R
-4s_2t_2R=0\label{coum33}\\ 
&&a_1 = -\frac {6s_3}c b_0 R - \frac{2s_2}c R,\quad\quad\quad a_2= 
- \frac {3s_3}c R^2\0\\
&&b_1 = -\frac {6t_3}c a_0 R - \frac{2t_2}c R,\quad\quad\quad b_2= 
- \frac {3t_3}c R^2\0
\b
From the previous section we see that we are indeed interested in finding 
all the solutions that have an analytic expansion in $x=n/N$ around 
$x=0$. In order to compute all possible solutions of this type, that is
the quantum vacua, we therefore drop the first term in the lhs of
(\ref{coum33}) and solve the resulting system. The third equation, in
particular, admits the solution $R=0$.
\a
&&R=0\label{R=0}\\
&& 3t_3 a_0^2+2t_2a_0 +cb_0=0\label{m33cl1}\\
&& 3s_3b_0^2+2s_2b_0 +c a_0=0\label{m33cl2}
\b
which give rise to four (in general) distinct solutions. The alternative
$R\neq 0$ leads to
\a
&&\frac 92 c \frac{s_3t_3}{s_2} a_0^2 + 3( \frac{s_3t_2}{s_2} c + 8s_2t_3)a_0 +
108\,s_3t_3a_0b_0+54 \frac{s_362t_2}{s_2}b_0^2 + 162
\frac{s_3^2t_3}{s_2}a_0b_0^2\0\\
&&~~~~~~~~~~~=3 \frac{s_3}{s_2}(c^2 -16 s_2t_2)+c^2-4s_2t_2, \label{Rnot01}\\ 
&&\frac 92 c \frac{s_3t_3}{t_2} b_0^2 + 3( \frac{t_3s_2}{t_2} c + 8s_3t_2)a_0 +
108\,s_3t_3a_0b_0+54 \frac{s_2t_3^2}{t_2}a_0^2 + 162
\frac{s_3 t_3^2}{t_2}a_0^2b_0\0\\
&&~~~~~~~~~~~=3 \frac{t_3}{t_2}(c^2 -16 s_2t_2)+c^2-4s_2t_2, \label{Rnot02} 
\b
This leads to an algebraic equation of order 10 for $a_0$, for instance. 
Therefore, generically, we have 10 (possibly complex) solutions for $a_0$, 
each of which 
gives rise to two different values for $b_0$. Altogether we 
are going to find 24 different quantum vacua. Once again it is interesting to 
compare these solutions with the classical ones. To this end in the above
equations we set $a_2=b_2=a_1=b_1=0$ as well as $R=0$, from which we get
eqs.({\ref{m33cl1},\ref{m33cl2}). 

From the first we can get $b_0= -\frac 1c ( 3t_3 a_0^2+2t_2a_0 )$, whence
we get either $a_0=0$ or the cubic equation
\a 
27 s_3t_3^2 a_0^3 + 36 t_2 s_3 t_3 a_0^2 +(12s_3t_2^2-6 c s_2t_3)a_0 +
c(c^2-4s_2t_2)=0\0
\b
Therefore in general we have four classical extrema. In ref.\cite{BBR} 
it is shown how to find an explicit series expansion in $x$ about each 
of these solutions.

Before we end this section it is interesting to discuss the geometric meaning 
of the first three equations in (\ref{coum33}). We can think of the third
equation as a definition of the complex $x=n/N$ plane. The two remaining
ones are quadratic equations in $a_0,b_0,R$. Introducing homogeneous 
coordinates they can be seen to represent two hypersurfaces in 
${\mathbb P}^3$. The intersection is a genus 1 Riemann surface.

\subsection{The ${\cal M}_{p_1,p_2}$ model}

In the general case the matrix rank for $Q(1)$ and $Q(2)$ was given above 
and the quantum EoMs become of course very complicated.
It is however simple to write down the equations that identify the 
extrema with classical analog. They are
\a
V_1'(a_0) + c b_0=0,\quad\quad\quad V'_2(b_0) +c a_0=0\label{extremaeq}
\b
while all the other fields are set to zero. We have in general 
$(p_1-1)(p_2-1)$ solutions of this type in perfect correspondence with the
classical analysis. The simplest solution is $a_0=b_0=0$. Other solutions may 
be hard or even impossible to determine explicitly. Anyhow, once one such
solution is known it is possible to find explicit expressions for the fields
around it in terms of $x=n/N$. The quantum Riemann surface corresponding
to this model has the same general structure as the one discussed at the 
end of the previous subsection. It is an intersection curve of two
hypersurfaces in ${\mathbb P}^3$. Its genus can be calculated with 
standard methods in  intersection theory. For instance, the quantum Riemann
surface for the ${\cal M}_{4,3}$ model is in general genus 2.

\section{Topological open string expansions}

\setcounter{equation}{0}
\setcounter{footnote}{0}

As we have pointed out in section 3 the matrix formalism allows us to compute
both closed and open string amplitudes. In this final section we would like
to show how to obtain the latter. To avoid clogging the text with cumbersome
formulas we work in the ${\cal M}_{2,2}$ model. The most explicit formulas 
for the latter can be found in \cite{BCX}. We adapt the results to the 
present situation
by rescaling all the coupling, such as $t_k\to t_k/g_s$, etc.
For instance the exact one point function is
\a
<\tau_r> = \sum_{2l=0}^r \sum_{k=0}^l \frac {r!2^{-k}}{(r-2l)! k!(l-k)!}
\left(\bac N \\ l-k+1\ea\right) g_s^l A^{l}  B^{r-2l} \label{taun}
\b
where
\a
A= \frac {2 s_2}{c^2-4s_2t_2 },\quad\quad B=\frac{2 s_2 t_1-cs_1}{c^2-4s_2t_2} 
\b
Now, for large $N$ it makes sense to expand the binomial coefficient in
(\ref{taun}) in powers of $N$: 
\a
\left(\bac N \\ s+1\ea\right) = \sum_{p=0}^{s} \frac {\beta_{s-p}(s)}{(s+1)!} 
(-1)^{p+s}\, N^p \label{betas}
\b
where 
\a
\beta_k(s) = \sum_{1\leq s_1 < s_2 \ldots < s_k \leq s}  
s_1s_2\ldots s_k,\qquad 1\leq k\leq s,\qquad \beta_0(s) =1,
\qquad\beta_k(s) = 0 \quad {\rm otherwise}\0
\b
As a consequence we get
\a
<\tau_r> = \sum_{2l=0}^r \sum_{k=0}^l \sum_{p=0}^{l-k+1} 
\frac {r!\, 2^{-k}\, (-1)^{p+l-k+1}}{(r-2l)! k!(l-k)!(l-k+1)!}
\beta_{l-k-p+1}(l-k) \,
 A^{l}\,  B^{r-2l}\, g_s^l\, N^p \label{taun1}
\b
Setting $p=h$ and $l=2g-2+h$, we can easily extract the contribution to the
correlator from world--sheets of genus $g$ with $h$ boundaries.
This of course requires $l-p$ to be even. But in (\ref{taun1}) we have 
also contributions with odd $l-p$. We do not have an interpretation for
these additional contributions. They may be perhaps related to the possibility
of describing the effect of punctures.

\vskip 1cm

{\bf Acknowledgments.}  This paper will appear in the proceedings of the
4th International Symposium "Quantum Theory and Symmetries" (QTS-4), Varna
(Bulgaria), August 2005, and of the International Workshop 
"Supersymmetries and Quantum Symmetries", Dubna (Russia), July 2005
(shortened version). This research was supported by the Italian MIUR
under the program ``Teoria dei Campi, Superstringhe e Gravit\`a''.
The work of G.B. is supported by
the Marie Curie European Reintegration Grant MERG-CT-2004-516466,
the European Commission RTN Program MRTN-CT-2004-005104.

\end{document}